\documentstyle[preprint,aps]{revtex}
\begin{document}
\title{Experimental evidence for a spin-Peierls-transition 
in $\alpha\prime-NaV_2O_5$}
\author{\bf M. Weiden$^a$, R. Hauptmann$^a$, C. Geibel$^a$, F. Steglich$^a$,\\ 
M. Fischer$^b$, P. Lemmens$^b$, G. G\"untherodt$^b$}
\address{$^a$ Technische Physik, TH Darmstadt, 
Hochschulstr. 8, 64289 Darmstadt, Germany \\
$^b$ 2. Physikalisches Institut, RWTH Aachen, 
Templergraben 55, 52056 Aachen, Germany}
\date{21.02.97}

\maketitle
key-words: spin-Peierls transition, magnetic susceptibility, 
raman scattering, \\
$\alpha\prime-NaV_2O_5$

\begin{abstract}
We present the first measurements of magnetisation and 
Raman light scattering on $\alpha\prime-NaV_2O_5$ single crystals. 
Below 34 K, we observe a 
pronounced isotropic decrease of the 
susceptibility indicating the opening of a spin gap. 
The transition temperature is slightly field 
dependent. Raman experiments reveal a crystallographic 
distortion at the transition. Our results 
clearly establish $\alpha\prime-NaV_2O_5$ to be the 
second inorganic spin-Peierls system. 
\end{abstract}
\newpage
\section{Introduction}
One of the unusual ground states encountered in 
quasi-one-dimensional spin-systems is the 
spin-Peierls (SP) groundstate, where a magneto-elastic 
coupling leads below the transition 
temperature $T_{SP}$ to a dimerisation of the 
spin-chain and thus to the formation of a non-
magnetic singlet \cite{Bul,Cro,Kho}. Up to 1993, this 
transition has only been observed in a few 
organic compounds \cite{Bra,Hui,Jac}. The discovery of a 
SP-state in the inorganic compound $CuGeO_3$ 
\cite{Has} has renewed strong interest in this phenomenon: 
for the first time, detailed studies of a SP-system 
on large single crystals \cite{Bou} as well as 
investigations of the effect of impurities \cite{Wei} 
on a SP-groundstate are possible. Yet, despite 
intensive research for other compounds, $CuGeO_3$ 
has remained the only known inorganic SP-system.
In 1996, Isobe et al. \cite{Iso} found a pronounced 
decrease of the susceptibility $\chi$ (T) in 
$\alpha\prime-NaV_2O_5$ below T $\approx$ 34 K and 
concluded the opening of a spin-gap. The 
remaining susceptibility below the transition 
temperature was attributed to Van-Vleck 
contributions as observed in $VO_2$, where a 
singulett is formed already above room 
temperature \cite{Goo,Kaw}. The orthorombic crystal 
structure \cite{Pou,Car} of $\alpha\prime-NaV_2O_5$ 
contains $V^{4+}$ (S =1/2)-chains along the
 b-axis, in which the spins are 
antiferromagnetically coupled with an exchange 
coupling J = 560 K \cite{Iso}. These chains are 
separated from each other by nonmagnetic 
$V^{5+}$-chains, which results in a quasi-one-
dimensional behaviour of the magnetic properties 
\cite{Car2,Mil}. 
Since Isobe et al. only performed 
measurements on polycrystalline samples, they 
were not able to study the orientation 
dependence of the magnetic transition itself 
and of the susceptiblity above the transition 
to proof a negligible influence of spin anisotropy 
and a primary isotropic Heisenberg-character of 
the spin-chains. Additionally, they were not able 
to confirm the existence of any 
crystallographic distortions at the transition 
temperature, a feature which is, however, essential 
for a SP-transition. Also, an analysis of the 
susceptibility on polycrystalline samples by Mila 
et al. \cite{Mil} revealed no transition at all despite a 
similar exchange coupling of J = 529 K.


\section{Experimental details and results}
The single crystals used for the investigations 
were grown by a self-flux method. Details of the 
preparation procedure will be published elsewhere \cite{Fis}. 
We observe a strong dependence of the transition on 
the Na-content: both excess and defficiency decrease 
the transition temperature. Susceptibility measurements 
were performed with a commercial SQUID-
magnetometer (MPMS, Quantum Design). 
Raman-measurements were performed with an 
Ar-Laser ($\lambda$= 514.5 nm) and a Dilor 
XY-spectrometer including a nitrogen-cooled 
CCD-detector.
The results of the measurements of the 
susceptibility $\chi$ (T) with magnetic field 
parallel to 
the three crystallographic axes are shown in 
fig. 1. As the crystals are thin platelets, 
several crystals were oriented parallel and 
fixed using a small amount of wax. This wax 
lead to a temperature-independent diamagnetic 
background, which size could not be determined 
precisely. Owing to the fact that at low 
temperatures $\chi$ (T) corresponds to a 
temperature-independent Van-Vleck-contribution, 
we shifted the experimental values to fit the 
value at T = 2 K obtained in a measurement without any 
wax. These shifts correspond to the addition 
of a temperature-independent value and will 
therefore neither affect the absolute size of 
the anomaly in $\chi$ (T) nor the temperature 
dependence. In contrast to the results of Isobe 
et al., our measurements show no Curie-contribution 
at low temperatures, indicating the high quality 
of the single crystals.
The decrease of $\chi$ (T) below the transition 
temperature presents the same shape and 
nearly the same magnitude along the three axes. 
Such an isotropic behaviour is the 
characteristic sign of the opening of a spin gap 
and cannot be due to another magnetic 
instability, e.g. a N\'e el-transition. The slight 
differences in the magnitude of $\chi$ (T) above 
the transition temperature can be accounted for by 
an anisotropy in the g-factor. Assuming 
g = 2 for B $\|$ c, this would lead to 
g = 2.06 for B $\|$ b and g = 2.14 for B 
$\|$ a. Apart for this difference, the 
temperature dependence of $\chi$ (T) is the same 
along the three axes indicating the isotropic 
Heisenberg-character of the spin-chains.
Below the transition temperature, the temperature 
dependence of the susceptibility was fitted 
using the theory of Bulaevskii \cite{Bul2}, where 
$\chi$ (T) of a dimerized chain is calculated 
using two parameters J and $\gamma$, J being 
the mean coupling constant, $\gamma$ describing 
the dimerization. The gap energy $\Delta$ can 
be deduced from J and $\gamma$ using the theories 
of Pytte \cite{Pyt} and Bray et al. \cite{Bra}. The result 
of the fit is shown in the inset of fig. 1. 
We obtain $\Delta \approx$ 85 1 15 K, J = 441 K; 
the values differ only slightly between the 
different directions and for similar analyses 
on polycrystals. J is in good agreement with 
J = 560 K \cite{Iso} or J = 529 K \cite{Mil} obtained from 
the high temperature susceptibility.
Investigations of the field dependence of the 
transition temperature of a polycrystalline sample 
revealed a small shift to lower temperatures 
$\Delta T \approx$ 0.15 K at B = 5.5 T.  The 
observed field dependence can be well described 
using the theory of Bulaevskii \cite{Bul2}and Cross
\cite{Cro} for the field dependence of a SP-system: 
$T_{SP}(0)/T_{SP}(B)-1=\alpha[(g\mu_BB)/(k_BT_{SP}(0))]^2$. 
We found $\alpha \approx$ 0.118, 
in good agreement with theoretical predictions of 
$\alpha$ = 0.091 - 0.11 for a SP-system \cite{Bul,Cro}. 
This field dependence clearly indicates that the 
transition is not a simple structural one.
Raman scattering experiments were performed in 
quasi-backscattering geometry. The 
polarization was parallel to the $V^{4+}$-chains (bb), 
perpendicular (aa) or crossed polarized 
(ab). Several phonons could be detected for 
temperatures above the transition temperature. 
Six modes were observed both in (aa) as well as 
in (bb) geometry: at 88, 178, 304, 420, 530 and 
971 cm$^{-1}$. Additionaly there is one mode at 
230 cm$^{-1}$ only observable in (aa) and 
one mode at 449 cm$^{-1}$ only observable in 
(bb)-polarization. In (ab) geometry, four 
modes were identified at 173, 262, 292 and 
685 cm$^{-1}$. Besides, small residual 
contributions could be observed at 449 and  
530 cm$^{-1}$. Along the chains (b-axis), the 
modes at 449 and 530 cm$^{-1}$ are strongly 
asymmetric and show a tail towards the high 
energy side. Taking into account that the energy 
of these modes are very close to the energy 
scale of spin excitations along the Heisenberg 
chains, we would suggest that these tails form 
due to magnetic Raman scattering. Further 
experiments are required to clarify the nature of 
this effect.
Below the transition temperature several new modes 
could be detected. In Fig. 2, spectra at 
T = 100 K and at T = 5 K in (bb) geometry are compared. 
Strong additional peaks are seen at 
65, 104, 130, 650 and 945 cm$^{-1}$. 
In addition, smaller contributions appear at 
230, 246 and 296 cm$^{-1}$. Also, a formerly 
symmetry forbidden mode from (ab)-polarization 
at 685 cm$^{-1}$ is now observable. As e.g. the 
mode at 945 cm$^{-1}$ can be asssigned to a 
phonon mode due to its lineshape, clear evidence 
is given that the transition leads to a 
symmetry breaking of the lattice.


\section{Conclusion and Summary}
We have grown for the first time single crystals of 
$\alpha\prime-NaV_2O_5$ and present the first 
investigations of the orientation-dependence of the 
susceptibility and the first Raman-results. 
We found at the transition temperature an isotropic 
decrease of the susceptibility pointing to 
the opening of a spin-gap and the formation of a 
spin-singlet. Also, the isotropy of the 
susceptibility proves the chains to be of isotropic 
Heisenberg-type. The Raman experiments 
indicate that this magnetic transition is accompanied 
by a crystallographic distortion. The field 
dependence of the transition temperature rules out a 
simple structural transition. 
The isotropic exponential disappearance of the 
spin-susceptibility, a simultaneous 
crystallographic distortion and the field dependence 
of the transition temperature are the 
hallmarks of a spin-Peierls transition. Therefore, 
our results clearly prove 
$\alpha\prime-NaV_2O_5$ to be the second inorganic SP-system, 
having the highest transition temperature 
T$_{SP}$ = 34 K of all known SP-systems.\\


Acknowledgment:
This work was supported by SFB 252, SFB 341 
and BMBF FKZ.13N6586 We acknowledge 
the help of P. Haberl in using the Ljour2-stylefile.



\begin{figure}
\caption{Susceptibility of $\alpha\prime-NaV_2O_5$ along 
the three major crystallographic axes. The 
values  for the magnetic field parallel to the three 
axes are shifted to reach the same value at 
T = 2 K (see text). At the transition temperature, an 
isotropic decrease is observed, 
which proves the opening of a spin gap. Inset: Analysis 
of the susceptibility of the 
single crystals for B $\|$ b using the theory of 
Bulaevskii. The solid line represents 
the fit with $\Delta \approx $ 85 K, J = 441 K.}
\end{figure}


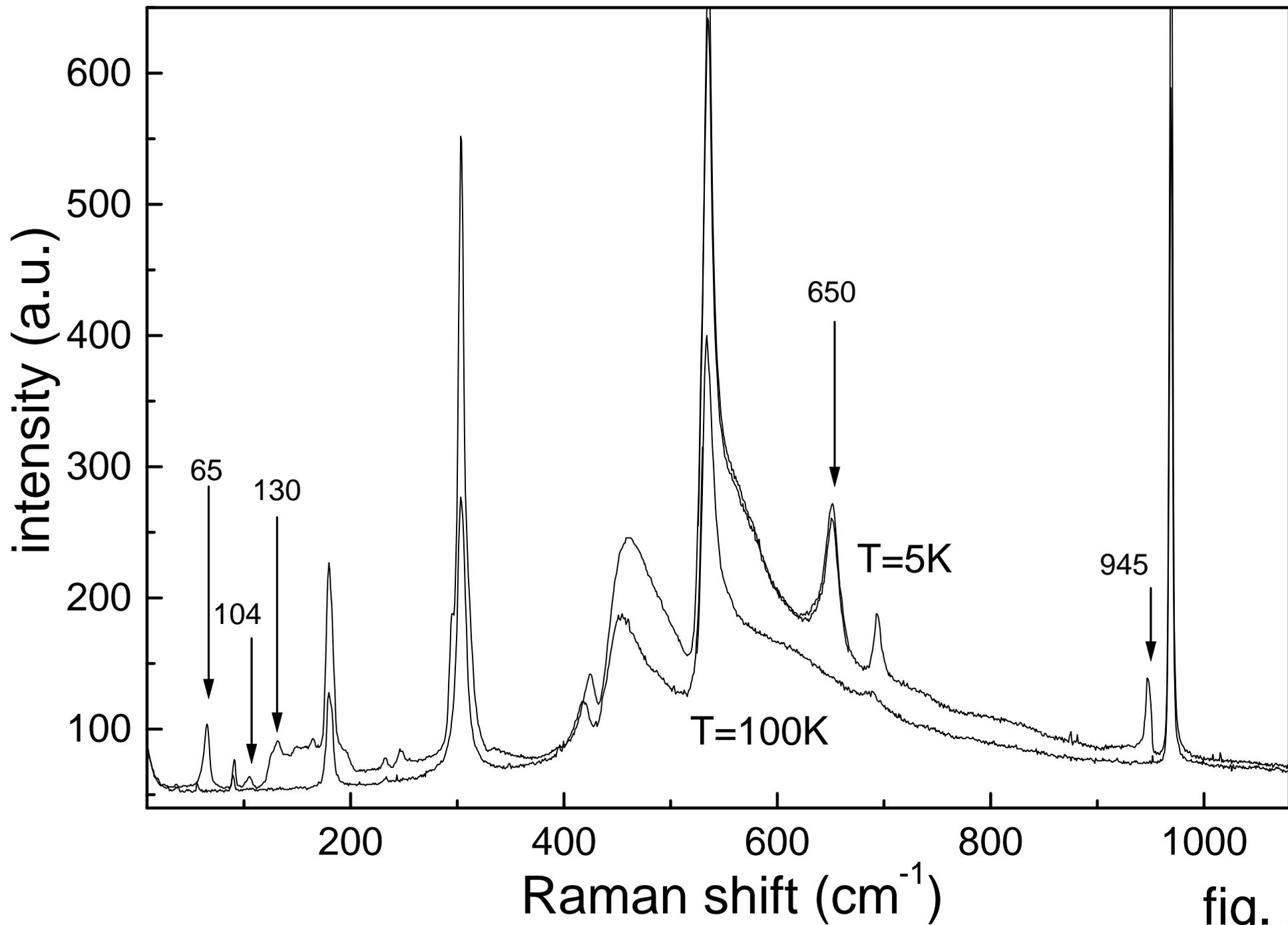
\begin{figure}
\caption{Ra\-man mea\-sure\-ments above and below the transition 
temperature in (b,b)-polarisation. Below the transition, 
several additional modes are observed.}
\end{figure}



\begin{thebibliography}{References}

\bibitem{Bul} Bulaevskii, L.N.: 
Soviet Phys. - Solid State 11, 921 (1969)

\bibitem{Cro} Cross, M.C.: Phys. Rev. B 20, 4606 (1979)

\bibitem{Kho} Khomskii, D., Geertsma, W., Mostovoy, M.: 
Czech. J. Phys. S5,  (1996)

\bibitem{Bra} Bray, J.W., Hart, H.R., Interrante, L.V.Jr., 
Jacobs, I.S., Kasper, J.S., Watkins, G.D., Wee, S.H., Bonner, J.C.: 
Phys. Rev. Lett. 35, 744 (1975)

\bibitem{Hui} Huizinga, S., Kommandeur, J., Sawatzky, G.A., 
Thole, B.T., Kopinga, K., deJonge, W.J.M., Roos, J.: 
Phys. Rev. B 19, 4723 (1979)

\bibitem{Jac} Jacobs, I.S., Bray, J.W., Hart, H.R., Interrante, L.V.Jr., 
Kasper, J.S., Watkins, G.D., Prober, D.E., Bonner, J.C.: 
Phys. Rev. B 14, 3036 (1976)

\bibitem{Has} M. Hase, I. Terasaki, K. Uchinokura, 
Phys. Rev. Lett. 70, 3651 (1993)

\bibitem{Bou} Boucher, J.P., Regnault, L.P.: 
submitted to J. de Phys.

\bibitem{Wei} M. Weiden, W. Richter, R. Hautmann, C. Geibel, 
F. Steglich, accepted by Physica B, and references therein 

\bibitem{Iso} M. Isobe, Y. Ueda, 
J. Phys. Soc. Jap. 65, 1178 (1996)

\bibitem{Goo} J.B. Goodenough, Progress in Solid State Chemistry 5, 
145, Pergamon Press New York (1971)

\bibitem{Kaw} T. Kawakubo, T. Nakagawa, 
J. Phys. Soc. Jap. 19, 517 (1964)       

\bibitem{Pou} M. Pouchard, A. Casalot, J. Galy, P. Hagenmuller, 
Bull. Soc. Chim. France 11, 4343 (1967)

\bibitem{Car} A. Carpy, J. Galy, Acta Cryst. B 31, 1481 (1975)

\bibitem{Car2} A. Carpy, A. Casalot, M. Pouchard, 
J. Galy, P. Hagenmuller, J. Sol.State Chem. 5, 229 (1972)

\bibitem{Mil} Mila, F., Millet, P., Bonvoisin, J.: 
preprint cond-mat/9605024 v2

\bibitem{Fis} M. Fischer, P. Lemmens, G. G\"untherodt, 
M. Weiden, R. Hauptmann, C. Geibel, F. Steglich, to be published

\bibitem{Bul2} Bulaevskii, L.N., Buzdin, A.I., Khomskii, D.I.: 
Sol. State Comm. 27, 5 (1978)

\bibitem{Pyt} Pytte, E.: Phys. Rev. B 10, 4637 (1974)

\end{thebibliography}
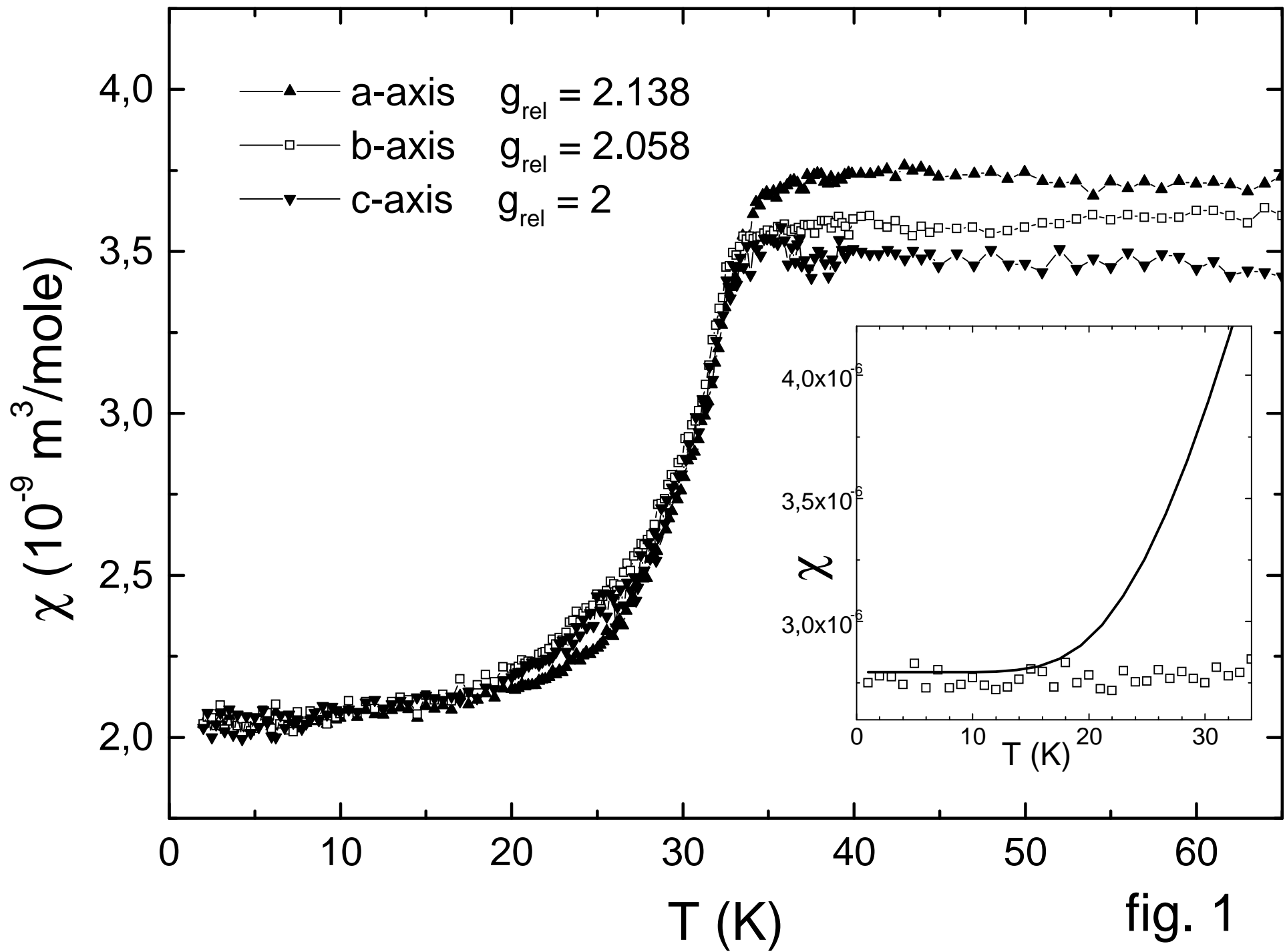
\end{document}